\documentclass[%
twocolumn,
english,
superscriptaddress,
floatfix,
reprint,
preprintnumbers,
 amsmath,amssymb,
 aps,
]{revtex4-2}

\usepackage{graphicx}
\usepackage{dcolumn}
\usepackage{bm}
\usepackage{hyperref}
\usepackage{cleveref}
\usepackage{color}
\usepackage{slashed}
\usepackage[dvipsnames]{xcolor}
\usepackage{booktabs}
\usepackage{multirow}
\usepackage{tikz}
\usetikzlibrary{decorations.markings}
\usepackage[qm]{qcircuit}
\usepackage{amsmath}
\usepackage[caption=false]{subfig}

\newcommand{\ket}[1]{|#1\rangle}

\definecolor{mygreen}{rgb}{0,0.5,0}
\definecolor{myblue}{rgb}{0,0,0.75}
\definecolor{mymagenta}{cmyk}{0,1,0,0.12}
\definecolor{mygray}{rgb}{0.5,0.5,0.5}

\graphicspath{{figures/}}

\begin{document}
\title{Physicality oracle for SU(3) Loop-String-Hadron dynamics: a digital quantum circuit}

\author{Fran Ilcic}
\email{p20230436@goa.bits-pilani.ac.in}
\email{ilcic.fran@gmail.com}
\affiliation{
    Department of Physics, BITS-Pilani, K K Birla Goa Campus, Zuarinagar, Goa 403726, India
}
\author{Indrakshi Raychowdhury}
 \email{indrakshir@goa.bits-pilani.ac.in}
\affiliation{
    Department of Physics, BITS-Pilani, K K Birla Goa Campus, Zuarinagar, Goa 403726, India
}
\affiliation{
    Center for Research in Quantum Information and Technology, Birla Institute of Technology and Science Pilani, K K Birla Goa Campus, Zuarinagar, Goa 403726, India
}

\begin{abstract}
Within the aim of understanding quantum chromodynamics through simulation, an increasingly studied approach is that of quantum computation and simulation. Challenges exist in encoding the minimal and physical degrees of freedom for a non-Abelian gauge theory and maintaining physical or gauge-invariant dynamics in a simulation. In this work, the Loop-String-Hadron (LSH) formulation of the 1+1-dimensional SU(3) lattice gauge theory is used to define an efficient mapping of SU(3) invariant degrees of freedom onto qubits. It is shown that the required number of qubits in the LSH basis is significantly reduced compared to its IRREP basis counterpart. While the non-Abelian Gauss laws of the SU(3) theory are automatically satisfied by the usage of LSH variables, the remnant constraints on the consistency of the flux numbers still exist. During time evolution, the noise can accumulate and take the state out of the sector of the Hilbert space where the constraint is satisfied. With this motivation, an oracle algorithm is constructed to be applied to the qubits for checking the local constraint at a given link. Costs in terms of qubit and gate number, and circuit depth, are found.
\end{abstract}
\maketitle

\date{\today}

\maketitle

\section{Introduction}

One of the frontiers in modern theoretical physics is that of understanding the dynamics of non-Abelian gauge field theories. This is primarily because one such theory, SU(3), is our best model of elementary nuclear particles and the strong force within the standard model \cite{SM,Gross:2022hyw}. While the theory itself is defined and has been widely successful, analytic or perturbative solutions cannot be obtained for all regimes. Consequently, understanding the dynamics and deriving predictions remain a significant challenge \cite{Gross:2022hyw}. One approach to getting closer to this goal has been \textit{lattice gauge theory} (LGT). By discretizing spacetime (looking at the continuum limit later) and thus reducing the degrees of freedom to a finite (or at least countable) number per unit volume, it has allowed for computer simulations of the gauge fields. Research on LGT has, since its beginnings in the 1970s \cite{LGT}, advanced in a few directions. The conventional method became the Monte Carlo simulation \cite{Creutz:1983ev,Wilson:1979wp}, and has been widely used since. Recently, new directions gained traction, including tensor networks computation \cite{rico2014tensor,tagliacozzo2014tensor,Meurice:2020pxc,mukunda1965tensor,Silvi:2014pta,Banuls:2018jag}. Hamiltonian approach, often focused on its diagonalization, is also used in some works \cite{Hamer:1981qt,Irving:1983sq,Umino:2000wh,Hamer:1984nj}. The work presented in this paper is done within the Hamiltonian formulation, which has the potential to provide efficient and insightful simulation, especially with quantum computers.

Traditional approaches to LGT often encounter a problem with real-time dynamics simulation due to a "sign problem" \cite{sign}. They also require exponentially scaled amounts of computing resources with increasing system size, as the system's quantum superposition involves entanglement of each basis state with any combination of other basis states. Hamiltonian LGT, originating with Kogut and Susskind \cite{KS}, uses unitary time evolution of the state defined by the Hamiltonian and avoids the sign problem \cite{LSHSU2,LSHSU3}. The formalism is further laid out in \cite{creutz1984quarks,smit2002introduction,zohar2015formulation}. Time evolution is more easily implementable and interpretable in terms of physical quantities, while the properties of the gauge theory's nature are retained. Here, time is real and separate from space, where the Hamiltonian is defined at any moment by the spatial configuration of the lattice. Fermions and the gauge field excitations are modeled by variables on the lattice's sites and links, respectively. The Hamiltonian is constructed as an operator acting on these variables, which in the continuum limit (the lattice spacing going to zero) produces the quantum field theory in question.

The idea of simulating physical theories on quantum computers originated and developed decades ago \cite{Benioff,Feynman,Lloyd} and has seen a renewed interest within the last decade due to the development of working (though not perfect by any stretch) quantum computers \cite{banuls2020review,Meurice:2020pxc,Bauer:2022hpo,DiMeglio:2023nsa,Bauer:2023qgm,Funcke:2023jbq}. Since quantum systems are probabilistic, probabilistic computers are more natural. Beyond the encoding of actual probabilities, quantum bits exist in superpositions of various states, making it possible to accurately simulate entangled states of physical systems. Ideally, with perfect quantum hardware, the computational advantage scales with the number of qubits, as the space of states grows exponentially compared to a classical computer. The number of variables grows with the dimension of the Hilbert space, which is exponential in system size, as there is a number for each superposition of the local quantum numbers. Qubits exist in superpositions themselves, so their required number scales linearly instead. In order to correctly represent theoretical quantum field theory models on lattices and then on quantum computers, appropriate formulations must be defined, and the qubit mapping must be made efficient. There have been many works in this area, developing LGTs and tailoring them for quantum simulation, both the Abelian \cite{Banerjee:2012pg,Kasper:2016mzj,zohar2017digital,rico2014tensor,muschik2017u1,zache2018quantum,klco2018quantum,schweizer2019floquet,davoudi2020towards,shaw2020quantum,Stetina:2020abi,Haase:2020kaj,Bauer:2021gek,Zhang:2021bjq,Ciavarella:2021lel,Huffman:2021gsi,Carena:2022kpg,Pardo:2022hrp,Davoudi:2024wyv,Farrell:2024fit,Farrell:2024mgu} and non-Abelian theories \cite{Chandrasekharan:1996ih,byrnes2006simulating,zohar2013cold,tagliacozzo2013simulation,banerjee2013atomic,wiese2013ultracold,wiese2014towards,tagliacozzo2014tensor,mezzacapo2015non,Banuls:2017ena,nuqs2019general,alexandru2019gluon,LSHSU2,Dasgupta:2020itb,klco2020su,Calajo,Atas:2021ext,Wiese:2021djl,Ciavarella:2021nmj,Davoudi:2022xmb,Atas:2022dqm,Ji:2022qvr,Mueller:2022xbg,Farrell:2022wyt,Farrell:2022vyh,Hartung:2022hoz,Alexandru:2023qzd,DAndrea:2023qnr,Romiti:2023hbd,Zache:2023dko,Kadam:2024zkj,Ciavarella:2024fzw}, with reviews \cite{dalmonte2016lattice,zohar2016quantum,banuls2020review,banuls2019simulating} and quantum hardware considerations \cite{martinez2016real,preskill2018quantum}.

Quantum simulation provides the extremely helpful benefit of parallel operation on all of the superposed states of the lattice, as the quantum computer's state also exists in a superposition. This is important because the Hamiltonian is not diagonal in the basis of states in which the positions of fermionic and field excitations on the lattice are defined. Hence, on-site quantum numbers can be encoded into the quantum hardware (qubits or qudits), and the Hamiltonian-generated evolution can be implemented through quantum gates. The input state of the lattice will quickly evolve into a large superposition of basis states, which would require exponential resources to simulate in a classical computation, but only linear in the quantum.

One of the main problems to be solved whenever doing lattice non-Abelian gauge theory is the insurance of gauge invariance and removal of unphysical states \cite{KS,Prepot}. Gauge invariant states form a subspace of the total Hilbert space spanned by the products of on-site and link variable spaces. When the state evolves in time by the Hamiltonian, gauge invariance is not broken, but since quantum computers are not perfectly coherent for unlimited periods of time, quantum simulated evolution will create errors in the form of contributions of states that (almost always) do break it. Gauge invariance is a global property of the state, but it is equivalent to a local law being satisfied all over the lattice. This local law is the (non-Abelian) Gauss law, which consists of all the chromoelectric fields being consistently parallelly transported across each lattice link.

To make the implementation as efficient as possible, \textit{Loop-String-Hadron} (LSH) \cite{LSHSU2,LSHSU3} formalism is used. It makes use of operators constructed in such a way that the basis states using the remaining degrees of freedom automatically solve the non-Abelian Gauss law constraints, which ensures the gauge invariance of the theory. Only a simple Abelian law remains to be checked for each link \cite{stryker2019oracles,LSHSU3}. This is the task that is solved, in the 1-dimensional case, by the oracle presented in this work. It uses an extra qubit for each link, in which it writes the information about whether the law is satisfied on that link. The essential principle is the same as in a previous work for the SU(2) case \cite{Gauss}.

In section \hyperref[sec:LSH]{II}, the Loop-String-Hadron formulation is summarized for the 1+1 dimensional lattice. The original Kogut-Susskind formalism (\hyperref[sec:KS]{II.A}) is reviewed first. Then the prepotential formulation (\hyperref[sec:Sch]{II.B}) is briefly shown in terms of Schwinger bosons, through which LSH can then be defined. For LSH itself (\hyperref[sec:KS]{II.C}), the basis is described as well as the non-Abelian Gauss law. In section \hyperref[sec:Qubit]{III}, the qubit cost is calculated with LSH applied and compared to the naive IRREP basis case. Section \hyperref[sec:oracle]{IV} explains the oracle in detail (\hyperref[sec:alg]{IV.A}) and provides component cost (\hyperref[sec:comp]{IV.B}) and circuit depth (\hyperref[sec:depth]{IV.C}) evaluation. Section \hyperref[sec:Qudit]{V} is a brief overview of \textit{qudits} - a multi-level analog of qubits, and the possibility and benefits they would provide for the system's encoding and the algorithm implementation. Closing remarks and future work directions are discussed in section \hyperref[sec:discus]{VI}.

\section{SU(3) lattice gauge theory coupled with fermions}
\label{sec:LSH}

\subsection{Kogut-Susskind formulation: constraints on the irrep basis}
\label{sec:KS}

Hamiltonian lattice gauge theory was first formulated by Kogut and Susskind \cite{KS}, and here a brief summary is given for the SU(3) case \cite{LSHSU3}. A lattice theory defines its variables (which are operators on the Hilbert spaces) on a lattice of vertices (sites) and on links connecting the nearest-neighbour sites. In our case, this is a 1-dimensional array of sites and links. The framework and the algorithm work for both open and periodic boundary conditions. A theory is defined by a Hamiltonian composed of lattice variables along with a group of gauge transformations under which it is symmetric. A Hamiltonian for a physical theory is built from the simplest possible gauge-invariant contractions of the operator variables of the entire lattice that reproduce the desired quantum field theory in the properly defined continuum limit (zero link length and infinite site density).

The gauge field is represented by operators on the links, defining the group elements that contract with the site variables. These group elements are the unitary holonomic operators $U^\alpha{}_\beta(r)$ acting as a transport of the gauge color across a link, transforming in the adjoint representation. The site variables are staggered fermion operators in the fundamental representation of the gauge group, $\psi^\alpha(r)$. To span the Hilbert space, usually a rigid rotator basis is considered. The chromoelectric field strength on the left and right side of each link $E^a(L,r)$, $E^a(R,r)$ (gauge invariant) act on the configuration space of a rigid rotator, which is equal to the sum of all irreducible representation spaces, thus spanning the gauge group space via the group Fourier transform.

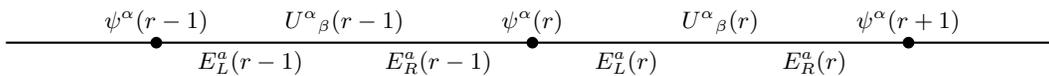
\begin{figure*}[]
    \centering
    \begin{tikzpicture}
        \foreach \x in {-1,0,1} {
            \filldraw (5*\x,0) circle (2pt);
        }
        
        \foreach \x in {-1,0} {
            \draw[thick] (5*\x,0) -- (5*\x+5,0);
        }
            \draw[thick] (-7,0) -- (-5,0);
            \draw[thick] (5,0) -- (7,0);

            \node[above] at (0,0) {$\psi^\alpha(r)$};
            \node[above] at (-5,0) {$\psi^\alpha(r-1)$};
            \node[above] at (5,0) {$\psi^\alpha(r+1)$};
        
            \node[above] at (-5+2.5,0) {$U^{\alpha}{}_{\beta}(r-1)$};
            \node[above] at (2.5,0) {$U^{\alpha}{}_{\beta}(r)$};
        
            \node[below] at (-5+1.25,0) {$E^a_L(r-1)$};
            \node[below] at (-5+3.75,0) {$E^a_R(r-1)$};
            \node[below] at (1.25,0) {$E^a_L(r)$};
            \node[below] at (3.75,0) {$E^a_R(r)$};
    \end{tikzpicture}
    \caption{Kogut-Susskind variables for a 1-dimensional lattice}
\end{figure*}

In the SU(3) case, this is a generalized rigid rotator, whereas the SU(2) theory gives an actual one. The rigid rotator's space of states is the space of all possible rotations. These can be thought of as matrices which, when acting on some space-fixed set of coordinate axes, produce the axes fixed to the rotator. Each rotation corresponds to an SU(2) element, and therefore to a state of the gauge field on a link. Gauge transformations are applied locally at the sites and thus to the adjacent ends of links. While the fermionic variables are transformed in the fundamental representation, the gauge field is transformed at each end of its link separately, so that the proper contraction of the site and link variables remains gauge-invariant. For the link, this is analogous to changing both the space-fixed and body-fixed axes in the rotator picture. These two are, in general, different, each corresponding to its own site's gauge transformation.

Continuing with the SU(2) case for illustrative purposes, the group Fourier transform takes the rotator to the angular momentum $j,m$ basis, where the Casimir $j$ is related to the chromoelectric field strength, which defines an irreducible representation of the group. This makes the space much simpler to work with as well as explicitly quantum. Any gauge group element acts on a vector in this basis as its version in the irreducible representation defined by $j$. As such, gauge transformations cannot change the value of $j$, but only that of $m$. It is therefore clear that for gauge invariant contractions the left and right Casimirs on any link need to be the same, while all the pairs of the projection number $m$ (one $m$ for each half of the link) need to be summed over. In the fundamental and anti-fundamental representations, this corresponds to the color indices on the fermions and the holonomic link operators. This is why the link operators have two color indices, which contract with the fermionic operators when forming a Hamiltonian. As for notation, the Greek letters denote the color indices $\alpha\in\{1,2,3\}$, while the Latin letters are used for adjoint indices $\mathrm{a}\in\{1,2,\dots,8\}$. Summation is implied with repeated indices. The argument $r$ denotes the position in the lattice and $s$ the side of a link. For fermionic variables, this means the coordinates of the site, and in the case of link variables, the direction in which the link leaves the site has to be specified additionaly.

The fermion operators obey the anticommutation relations
\begin{equation}
\begin{aligned}
& \left\{\psi^{\alpha}(r), \psi_{\beta}^{\dagger}\left(r^{\prime}\right)\right\}=\delta_{\beta}^{\alpha} \delta_{r r^{\prime}},
\end{aligned}
\end{equation}
with all other combinations equal to zero; whereas the link operators obey the relations
\begin{equation}
\begin{aligned}
{\left[E^{\mathrm{a}}(s, r), E^{\mathrm{b}}\left(s^{\prime}, r^{\prime}\right)\right] } & =\delta_{s s^{\prime}}\delta_{r r^{\prime}} i f^{\mathrm{abc}} E^{\mathrm{c}}(s, r),
\end{aligned}
\end{equation}
\begin{equation}
\begin{aligned}
{\left[U^{\alpha}{}_{\beta}(r), U^{\gamma}{}_{\eta}\left(r^{\prime}\right)\right] } & =\left[U^{\alpha}{}_{\beta}(r), U^{\dagger \gamma}{ }_{\eta}\left(r^{\prime}\right)\right]=0,
\end{aligned}
\end{equation}
and their commutators with each other make the holonomic operators $U^\alpha{}_\beta(r)$ the SU(3) variant of ladder operators for the chromoelectric fields:
\begin{equation}
\begin{aligned}
{\left[E^{\mathrm{a}}(L, r), U\left(r^{\prime}\right)\right] } & =-\delta_{r r^{\prime}} T^{\mathrm{a}} U(r),\\
{\left[E^{\mathrm{a}}(R, r), U\left(r^{\prime}\right)\right] } & =\delta_{r r^{\prime}} U(r) T^{\mathrm{a}}.
\end{aligned}
\end{equation}
Here $f^{abc}$ are the structure constants and $T^a$ the generators (in the fundamental representation) of $SU(3)$. The letter $s$ labels the side of the link and can take the values left ($L$) or right ($R$).

This defines a Hilbert space spanned by the fermion quantum numbers and the rigid-rotator numbers that define the link state. However, gauge invariance still needs to be imposed. In $dim$ spatial dimensions, the local gauge transformations are generated by the operators
\begin{equation}
\begin{aligned}
G^{\mathrm{a}}(r)=&\sum_{i=1}^{dim}\big(E^{\mathrm{a}}(L, r,i)+E^{\mathrm{a}}(R,r,-i)\big)+\\
&+\psi^{\dagger}(r)T^{\mathrm{a}}\psi(r),
\end{aligned}
\end{equation}
which in $dim=1$ reduces to
\begin{equation}
\begin{aligned}
G^{\mathrm{a}}(r)=&E^{\mathrm{a}}(L,r)+E^{\mathrm{a}}(R,r-1)+\\
&+\psi^{\dagger}(r)T^{\mathrm{a}}\psi(r),
\end{aligned}
\end{equation}
This means that applying the gauge transformation operators generated by $G^{\mathrm{a}}(r)$ to a state effectively turns it into an equivalent state for which the physics is the same. This translates into the requirement that the Hamiltonian commute with all gauge transformations, making all dynamics equivalent for such states:
\begin{equation}\label{commutation}
[H,G^{\mathrm{a}}(r)]=0.
\end{equation}
If one now combines all equivalent states in an appropriate way, one can obtain a state that transforms into itself. Consequently, a sector of the Hilbert space can be chosen that is unaffected by the transformations. In terms of the generators, this means
\begin{equation}
G^{\mathrm{a}}(r)|\Psi\rangle=0 \quad \forall \mathrm{a}, r.
\end{equation}
These constraints are called Gauss laws. Due to (\ref{commutation}), time evolution will never take a state from this sector out of it. States inside it are called gauge-invariant, and can be gotten from the vacuum via gauge-invariant operators. These are constructed by contracting all of the color indices in arbitrary products of site and link operator variables, effectively creating loops, strings, or hadrons on the lattice. Each contracted pair of indices must be local to a single lattice site, thus connecting the transformation of the gauge field and the adjacent fermion, which indeed transform with the same group element. The Hamiltonian is designed to be gauge-invariant and have the continuum limit equal to the Hamiltonian of the QCD quantum field theory. The exact form of the Hamiltonian is not relevant to this paper, but it is worth noting that the links' holonomic operators contracted with adjacent fermions give the hopping terms that control the dynamics.

To summarize, a local space of a link can be spanned by a basis of irreducible representations. Since all of the color indices are contracted, the gauge-invariant states can only have an IRREP defined and not any state within the representation space. This is easy to visualize in the context of SU(2), where a single $j$ value defines an IRREP, while its basis elements are $\ket{j,m}$. The $j$ value denotes the representation, and the magnetic number $m$ specifies a basis state within it. The gauge field $U(r)$ of a link is equivalent to a 3D rigid body rotation from one orientation of the axes to another. A local gauge transformation $V(r)$ acts differently at the two ends of the link, changing the link's gauge field as $U'(r)=V(L,r)U(r)V^{-1}(R,r)$. This means that the rigid body's initial and final orientations are rotated separately. However, these rotations cannot take the state out of its IRREP. In other words, the $j$ number of the two ends of a link must be equal. The link's configuration can thus be specified by the $\ket{j,m_L,m_R}$ ket (as obtained by the group Fourier transform), which can be thought of as a rotation from a 3D space-fixed set of axes to the axes of the body's frame, as discussed earlier. Gauge transformations then separately rotate each frame, changing $m_L$ and $m_R$. In this way, one can choose the configuration within the IRREP by setting the gauge, but the IRREP itself cannot be changed, and is thus physical.

\begin{figure}[h]
    \centering
    \begin{tikzpicture}
        \foreach \x in {0,1} {
            \filldraw (4*\x,0) circle (2pt);
        }
        
            \draw[thick] (0,0) -- (4,0);
            \draw[thick] (-1,0) -- (0,0);
            \draw[thick] (4,0) -- (5,0);
        
            \node[above] at (2,0) {$j$};
        
            \node[below] at (1,0) {$m_L$};
            \node[below] at (3,0) {$m_R$};
    \end{tikzpicture}
    \caption{Kogut-Susskind SU(2) rigid rotator variables}
\end{figure}
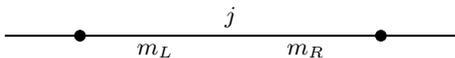

To form gauge-invariant states, the operator products must be summed over all $m$ numbers, giving superpositions that are invariant under rotations at each link end. This effectively leaves only the $j$ quantum number as an actual variable. It describes the amount of chromoelectric flux passing through the link. In other words, the group indices at all sites and link ends have to be contracted to form gauge singlets. If using only link operators, this can be done at all of the involved sites only if the links used in forming the singlet close into a loop. Otherwise, the beginning and the ending of the array of links would be sites for which one of the indices has no link to contract with. The resulting space of gauge invariant states is a superposition of all possible closed loops on the lattice, each having a defined $j$ quantum number. If site variables are considered as well, strings and hadrons are allowed too, as the fermions can contract with the final links or with themselves. In one spatial dimension, loops only exist as flux passing through the entire lattice.

The above construction is known as the rigid rotator basis. In SU(3) theory, its analog has two quantum numbers denoting each IRREP, since the group is rank two. Equivalently, it has two independent types of chromoelectric flux. In usual notation, there are three more numbers per link side that define the starting and ending state within the IRREP. The Hilbert space basis vectors for each link are therefore $\ket{p,q;I_L,I^z_L,Y_L,I_R,I^z_R,Y_R}$. In this basis, just like with SU(2), gauge invariance is achieved by combining the local operators into loops, strings, or hadrons. Such combinations will turn gauge invariant states from one into another, and by acting on the vacuum they span the physical states. But the local basis vectors can be combined into global states in other ways, too. This means that they span the whole Hilbert space, and not just the gauge-invariant part, which contains all of the physics. For quantum simulations, this is undesirable, as it will require a lot of qubits for encoding and has a lot of room for error from noise. Additionally, the Gauss law constraints themselves are non-Abelian, meaning they do not commute with each other. This is why the Loop-String-Hadron framework is favourable, as will be shown.

\subsection{Schwinger bosons or Prepotentials: local gauge invariance}
\label{sec:Sch}

While it is true that the original variables come from making analogies with the continuous QFT case, one can always undergo a change of variables without changing the model \cite{raychowdhury2019low,anishetty2018mass}. The prepotential formulation \cite{mathur2005harmonic,mathur2006loop,mathur2007loop,anishetty2010prepotential,raychowdhury2013prepotential} uses this to reformulate the theory in terms of variables that are all local to the lattice sites.

The insight of Schwinger bosons in the SU(2) case is that the Hilbert space of two simple harmonic oscillators is equivalent to the space of all SU(2) group elements. This is easily seen in the angular momentum basis. A state $\ket{j,m}$ is mapped to a state of the two bosons with $2j$ excitations in total, and a difference between the numbers of excitations in one versus the other equal to $2m$. The two harmonic oscillators have number operators $N_{1,2}=a_{1,2}^{\dagger}a_{1,2}$, where the ladder operators are defined by
\begin{equation}
[a_1,a_1^{\dagger}]=1=[a_2,a_2^{\dagger}],
\end{equation}
with other commutators giving zero. Now their states can be written as $\ket{N_1,N_2}$, where each of the ladder operators acts only on its half of the state. Using these, one can define operators
\begin{equation}
\begin{aligned}
J_+&=a_1^\dagger a_2,\\
J_+&=a_2^\dagger a_1,\\
J_z=\frac{1}{2}(&a_1^\dagger a_1-a_2^\dagger a_2),
\end{aligned}
\end{equation}
which can be shown to satisfy the usual angular momentum algebra (with $\hbar=1$). This leads to
\begin{equation}
J_z\ket{N_1,N_2}=\frac{1}{2}(N_1-N_2)\ket{N_1,N_2},
\end{equation}
which, for a fixed value of $j:=\frac{1}{2}(N_1+N_2)$, is a number between $-j$ and $j$ in steps of $1$. Thus $m:=\frac{1}{2}(N_1-N_2)$, and the angular momentum states $\ket{j,m}$ are attained.

\begin{figure}[h]
    \centering
    \begin{tikzpicture}
        \foreach \x in {0,1} {
            \filldraw (6*\x,0) circle (2pt);
        }
        
            \draw[thick] (0,0) -- (6,0);
            \draw[thick] (-1,0) -- (0,0);
            \draw[thick] (6,0) -- (7,0);
        
            \node[below] at (1.5,0) {$a_1(L,r),a_2(L,r)$};
            \node[below] at (4.5,0) {$a_1(R,r),a_2(R,r)$};
    \end{tikzpicture}
    \caption{SU(2) Schwinger bosons - variables are now only on the link ends, with none on the entire link itself.}
\end{figure}
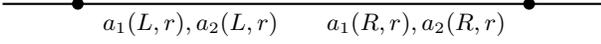

For SU(3) the same logic is used, but since it is a rank 2 group with a 3-dimensional fundamental representation, 2 sets of 3 harmonic oscillators are used, $a^{\alpha}(s, r),$ $b_{\alpha}(s, r)$. The positions of the indices here denote the representation with which the gauge transformations are applied - either fundamental or antifundamental. They have to contract with each other to form an invariant operator. One set of the six ladder operators exists for every end of every link, and they satisfy
\begin{equation}
\begin{aligned}
& {\big[a^{\alpha}(s, r), a^{\alpha^{\prime}}\left(s^{\prime}, r^{\prime}\right)\big]=\big[b_{\alpha}(s, r), b_{\alpha^{\prime}}\left(s^{\prime}, r^{\prime}\right)\big]=0},\\
& {\big[a^{\alpha}(s, r), b_{\alpha^{\prime}}\left(s^{\prime}, r^{\prime}\right)\big]=\big[a^{\alpha}(s, r), b^{\dagger \alpha^{\prime}}\left(s^{\prime}, r^{\prime}\right)\big]=0},\\
& \big[a^{\alpha}(s, r), a_{\alpha^{\prime}}^{\dagger}\left(s^{\prime}, r^{\prime}\right)\big]=\delta_{\alpha^{\prime}}^{\alpha} \delta_{s s^{\prime}} \delta_{r r^{\prime}},\\
&\big[b_{\alpha^{\prime}}(s, r), b^{\dagger \alpha}\left(s^{\prime}, r^{\prime}\right)\big]=\delta_{\alpha^{\prime}}^{\alpha} \delta_{s s^{\prime}} \delta_{r r^{\prime}},
\end{aligned}
\end{equation}
and $a^{\alpha}(s, r)\ket{0,0}=b_{\alpha}(s, r)\ket{0,0}=0$ (here the zeroes labeling the state are the Casimir numbers of the IRREPs).

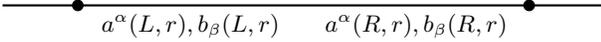
\begin{figure}[h]
    \centering
    \begin{tikzpicture}
        \foreach \x in {0,1} {
            \filldraw (6*\x,0) circle (2pt);
        }
        
            \draw[thick] (0,0) -- (6,0);
            \draw[thick] (-1,0) -- (0,0);
            \draw[thick] (6,0) -- (7,0);
        
            \node[below] at (1.5,0) {$a^\alpha(L,r),b_\beta(L,r)$};
            \node[below] at (4.5,0) {$a^\alpha(R,r),b_\beta(R,r)$};
    \end{tikzpicture}
    \caption{SU(3) Schwinger bosons. There are two triplets at each site as there is a fundamental and an anti-fundamental representation, each of dimension three, at every link end. These are not the proper IRREP degrees of freedom, but their appropriate combinations provide them.}
\end{figure}

As explained in \cite{LSHSU3}, these operators do not form irreducible representations when naively applied to the vacuum. However, irreducible versions $A^{\dagger}_{\alpha}(s, r)$ and $B^{\dagger\alpha}(s, r)$ can be constructed by adding terms with contractions $a^{\dagger} \cdot b^{\dagger}$ \cite{LSHSU3}.
\begin{equation}
\begin{aligned}
A_{\alpha}^{\dagger} & \equiv a_{\alpha}^{\dagger}-\frac{1}{\hat{N}_{a}+\hat{N}_{b}+1}\left(a^{\dagger} \cdot b^{\dagger}\right) b_{\alpha}\\
B^{\dagger \alpha} & \equiv b^{\dagger \alpha}-\frac{1}{\hat{N}_{a}+\hat{N}_{b}+1}\left(a^{\dagger} \cdot b^{\dagger}\right) a^{\alpha},
\end{aligned}
\end{equation}
with the Schwinger boson number operators
\begin{equation}
\hat{N}_{a} \equiv a^{\dagger} \cdot a=\sum_{\alpha=1}^{3} a_{\alpha}^{\dagger} a^{\alpha}\;\; \textrm{and}\;\; \hat{N}_{b} \equiv b^{\dagger} \cdot b=\sum_{\beta=1}^{3} b^{\dagger \beta} b_{\beta}.
\end{equation}
$A^{\alpha}(s, r)$ and $B_{\alpha}(s, r)$ directly produce irreducible representation states from the vacuum,
\begin{equation}\label{irrepstates}
\ket{P,Q}_{\vec\alpha}^{\vec\beta}=\mathcal NA_{\alpha_1}^\dagger\dots A_{\alpha_P}^\dagger B^{\dagger\beta_1}\dots B^{\dagger\beta_Q}\ket{0,0},\\
\end{equation}
at the expense of some of the commutation relations becoming more complicated ($\simeq$ denotes the equality on the subspace spanned by the $\ket{P,Q}_{\vec\alpha}^{\vec\beta}$ IRREPs above),
\begin{equation}
\begin{aligned}
{\left[A^{\alpha}, A_{\beta}^{\dagger}\right] } & \simeq\left(\delta_{\beta}^{\alpha}-\frac{1}{\hat{N}_{a}+\hat{N}_{b}+2} B^{\dagger \alpha} B_{\beta}\right),\\
{\left[B_{\alpha}, B^{\dagger \beta}\right] } & \simeq\left(\delta_{\alpha}^{\beta}-\frac{1}{\hat{N}_{a}+\hat{N}_{b}+2} A_{\alpha}^{\dagger} A^{\beta}\right),\\
{\left[A^{\alpha}, B^{\dagger \beta}\right] } & \simeq-\frac{1}{\hat{N}_{a}+\hat{N}_{b}+2} B^{\dagger \alpha} A^{\beta},\\
{\left[B_{\alpha}, A_{\beta}^{\dagger}\right] } & \simeq-\frac{1}{\hat{N}_{a}+\hat{N}_{b}+2} A_{\alpha}^{\dagger} B_{\beta},
\end{aligned}
\end{equation}
Here $\mathcal N$ is a normalization constant. Quantum numbers $\hat N_A=A_\alpha^\dagger A^\alpha$ and $\hat N_B=B_\alpha^\dagger B^\alpha$ are equal to ${N}_{a}$ and ${N}_{b}$ on states constructed by (\ref{irrepstates}). Values of indices in (\ref{irrepstates}) also define the state within the representation, usually denoted with the quantum numbers $I,I^z,Y$ \cite{mukunda1965tensor,Chaturvedi:2002si}. The chromoelectric fields can also be expressed in terms of these irreducible Schwinger bosons \cite{LSHSU3},
\begin{equation}
\begin{aligned}
E^{\mathrm{a}}(L, r)= & A_{\alpha}^{\dagger}(L, r)\left(T^{\mathrm{a}}\right)^{\alpha}{ }_{\beta} A^{\beta}(L, r) \\
& -B^{\dagger \alpha}(L, r)\left(T^{* \mathrm{a}}\right)_{\alpha}{ }^{\beta} B_{\beta}(L, r) ,\\
E^{\mathrm{a}}(R, r-1)= & A_{\alpha}^{\dagger}(R, r)\left(T^{\mathrm{a}}\right)^{\alpha}{ }_{\beta} A^{\beta}(R, r) \\
& -B^{\dagger \alpha}(R, r)\left(T^{* \mathrm{a}}\right)_{\alpha}{ }^{\beta} B_{\beta}(R, r) .
\end{aligned}
\end{equation}
Note that the prepotentials on the right side of a link are labeled by the site they are adjacent to, not by the one to their left, unlike the chromoelectric fields. One could proceed to use the number operators $\hat N_A$ and $\hat N_B$, but it turns out to be very useful to first combine them with fermionic variables into specific gauge-invariant products that define the building blocks of the Loop-String-Hadron formulation. Here, the main goal of using the formulation is to simplify the quantum numbers as much as possible, and solve the Gauss laws implicitly.

\subsection{LSH basis in 1+1 dimension}
\label{sec:LSHsub}

By looking at the irreducible representation basis \cite{KS}, the gauge field's degrees of freedom of a link were expressed in the basis of transformations, each of which takes place between two states within a single IRREP. With Schwinger bosons, these states are considered separately, each as a state of its own link end. By doing this, one introduces the requirement that the IRREP should be the same for the two ends of the link, since new degrees of freedom would otherwise be added to the original theory. This comes down to the Abelian Gauss law constraints \cite{LSHSU3}
\begin{equation}
\begin{aligned}
& \left(\hat{N}_{A}(L, r)-\hat{N}_{B}(R, r+1)\right)|\Psi\rangle \\
& =\left(\hat{N}_{B}(L, r)-\hat{N}_{A}(R, r+1)\right)|\Psi\rangle=0
\end{aligned}
\end{equation}
Both link ends still need to have their own states within the IRREP, as these are affected by gauge transformations separately. Now the focus can be shifted to each lattice site, at which the fermion content and the Schwinger bosons from both sides are transformed with the same gauge group element. This is an important step - turning to exclusively on-site variables. It allows us to eventually build the Hilbert space as the tensor product of single-site spaces with Abelian Gauss law constraints.

Looking at a single site, gauge invariance is ensured by the exclusive use of operators that have all their SU(3) indices contracted. The same principle is used in the original Kogut-Susskind formulation, but now there are only on-site operator contractions. For example,
\begin{equation}
\begin{aligned}
&\psi^\dagger_\alpha(r)B^{\dagger\alpha}(L,r),\quad \psi^\dagger_\alpha(r)A^\alpha(L,r),\\
&\psi^\dagger_\alpha(r)\epsilon^{\alpha\beta\gamma}A^\dagger_\beta(L,r)B_\gamma(L,r)
\end{aligned}
\end{equation}
are such on-site SU(3) singlets. To simplify notation, now that all operators are on-site, instead of naming the two half-links around a site $(R,r-1)$ and $(L,r)$, we call the two directions from a site $(\underline1,r)$ and $(1,r)$. This convention becomes even more justified when the system has more dimensions, where directions $\underline 2$ and $2$, $\underline 3$ and $3$ etc. also exist. In this notation, the above operators are written $\psi^\dagger\cdot B^\dagger(1),\;\psi^\dagger\cdot A(1),\;\psi^\dagger\cdot A^\dagger(1)\wedge B(1)$, where we also used the scalar product, $\cdot$, and the cross product, $\wedge$. Another set can be formed by using the $\underline1$ versions of the bosons instead of $1$.

Parts of the Hamiltonian such as $\psi^\dagger_\alpha(r)U^\alpha_{\;\,\beta(r)}\psi^\beta(r+1)$ can be constructed as products of separate contractions at sites $r$ and $r+1$ \cite{LSHSU3}.

Using all of these contractions, one gets a formulation in terms of combinations of fermionic and bosonic operators that change the content of a site and the flux coming out of it in a diagrammatic way (Figure 4 in \cite{LSHSU3}), or act as number operators. For example, an operator may create one of the three kinds of fermions at a site, and along with it a unit of one type of flux emerging from that site into one direction. Other operators would create only a flux line passing through the site, or a hadron without emerging flux, etc. More importantly, they define an orthonormal basis that is much more efficient in terms of free variables than the original one, allowing for efficient qubit digitization. This will be shown in the next section.

\begin{figure}[h]
    \centering
    \begin{tikzpicture}
        \draw[rounded corners, thick] (-2, -1.55) rectangle (2, 4);
        
        \node at (0, 4.2) {Site $r$};
        
        \node[below right] at (-2, 4) {$\;\nu_{\underline 1} \;\; \nu_0 \;\; \nu_1$};
        
        \foreach \i/\j/\k [count=\n] in {0/0/0, 0/0/1, 0/1/0, 0/1/1, 1/0/0, 1/0/1, 1/1/0, 1/1/1} {
            \node[right] at (-2, {3.4 - 0.5*(\n-1)}) {$\;\,$\i$ \quad $\j$ \quad $\k};
            
            \begin{scope}[shift={(0.7, {3.4 - 0.5*(\n-1)})}]
                \ifnum\i=1
                    \filldraw (-0.5,0) circle (2.5pt);
                \else
                    \draw (-0.5,0) circle (2.5pt);
                \fi
                \ifnum\j=1
                    \filldraw (0,0) circle (2.5pt);
                \else
                    \draw (0,0) circle (2.5pt);
                \fi
                \ifnum\k=1
                    \filldraw (0.5,0) circle (2.5pt);
                \else
                    \draw (0.5,0) circle (2.5pt);
                \fi
            \end{scope}

            \draw[orange, thick, postaction={decorate},decoration={markings, mark=at position 0.75 with {\arrow{stealth};}}] (1.2,2.9) -- (1.7,2.9);
            \draw[orange, thick, postaction={decorate},decoration={markings, mark=at position 0.3 with {\arrow{stealth};}}] (-0.3,2.4) -- (0.7,2.4);
            \draw[blue, thick, postaction={decorate},decoration={markings, mark=at position 0.85 with {\arrow{stealth reversed};}}] (0.7,2.4) -- (1.7,2.4);
            \draw[blue, thick, postaction={decorate},decoration={markings, mark=at position 0.7 with {\arrow{stealth reversed};}}] (1.2,1.9) -- (1.7,1.9);
            \draw[blue, thick, postaction={decorate},decoration={markings, mark=at position 0.5 with {\arrow{stealth reversed};}}] (-0.3,1.4) -- (0.2,1.4);
            \draw[blue, thick, postaction={decorate},decoration={markings, mark=at position 0.5 with {\arrow{stealth reversed};}}] (-0.3,0.9) -- (0.2,0.9);
            \draw[orange, thick, postaction={decorate},decoration={markings, mark=at position 0.75 with {\arrow{stealth};}}] (1.2,0.9) -- (1.7,0.9);
            \draw[orange, thick, postaction={decorate},decoration={markings, mark=at position 0.6 with {\arrow{stealth};}}] (-0.3,0.4) -- (0.2,0.4);

        }
        \node[right] at (-2, -0.8) {$\;\;n_P=2$};
        \draw[orange, thick, postaction={decorate},decoration={markings, mark=at position 0.53 with {\arrow{stealth};}}] (-0.3,-0.73) -- (1.7,-0.73);
        \draw[orange, thick, postaction={decorate},decoration={markings, mark=at position 0.53 with {\arrow{stealth};}}] (-0.3,-0.87) -- (1.7,-0.87);
        \node[right] at (-2, -1.3) {$\;\;n_Q=3$};
        \draw[blue, thick, postaction={decorate},decoration={markings, mark=at position 0.51 with {\arrow{stealth reversed};}}] (-0.3,-1.16) -- (1.7,-1.16);
        \draw[blue, thick, postaction={decorate},decoration={markings, mark=at position 0.51 with {\arrow{stealth reversed};}}] (-0.3,-1.3) -- (1.7,-1.3);
        \draw[blue, thick, postaction={decorate},decoration={markings, mark=at position 0.51 with {\arrow{stealth reversed};}}] (-0.3,-1.44) -- (1.7,-1.44);
    \end{tikzpicture}
    \caption{Illustration of how the two kinds of flux, $P$ (arrow to the right) and $Q$ (arrows to the left), can be generated on either side of the site by certain combinations of fermionic numbers $\nu_{\underline 1}$, $\nu_0$ and $\nu_1$. The bosonic quantum numbers $n_P$ and $n_Q$ define the number of flux lines that pass through the site and get added to the ones generated at the site by the fermions. As an example they are given as 2 and 3 here.}
    \label{LSHrules}
\end{figure}
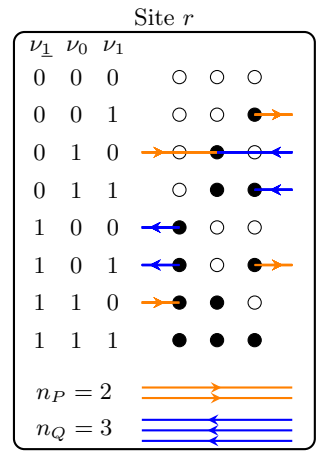

 The local Hilbert space is spanned by an orthonormal basis of the form
\begin{equation}
\ket{n_P,n_Q;\nu_{\underline 1},\nu_0,\nu_1}_r.
\end{equation}
The three kinds of fermions are named $\underline 1$, $0$, and $1$. The bosonic quantum numbers $n_P$ and $n_Q$ can take values of any nonnegative integers and represent the two kinds of chromoelectric flux passing through the site $r$ (loops). The three new fermionic variables $\nu_i$ determine both the fermionic content of the site (including hadrons) as well as the type $(P$ or $Q)$ and direction $(\underline 1$ or $1)$ of any flux lines that might be emerging from the site in either direction (strings). Thus, the LSH degrees of freedom can easily be visualized as rules for the presence of string ends at each site (based on fermion numbers), and two types of flux lines traversing the lattice until another such end terminates each. This is illustrated in Figure \ref{LSHrules}. For each of these quantum numbers, there is a gauge-invariant number operator (named the same as the quantum number itself). The details of this construction are given in \cite{LSHSU3}. Along with the combined raising and lowering operators, the LSH Hamiltonian can be constructed from these quantum numbers. Ultimately, the goal is to simulate on quantum computers the evolution of the lattice under this kind of Hamiltonian, which, when all types of fermions are included, should be equivalent to the original discretized QCD.

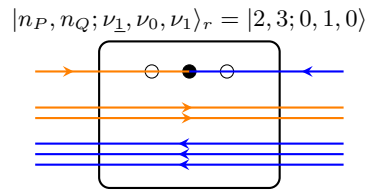
\begin{figure}[h]
    \centering
    \begin{tikzpicture}
        \draw[rounded corners, thick] (-1.2, -1.55) rectangle (1.2, 0.4);
        
        \node[above] at (0, 0.4) {$\ket{n_P,n_Q;\nu_{\underline 1},\nu_0,\nu_1}_r=\ket{2,3;0,1,0}$};

        \draw (-0.5,0) circle (2.5pt);
        \filldraw (0,0) circle (2.5pt);
        \draw (0.5,0) circle (2.5pt);

        \draw[orange, thick, postaction={decorate},decoration={markings, mark=at position 0.25 with {\arrow{stealth};}}] (-2.05,0) -- (0,0);
        \draw[blue, thick, postaction={decorate},decoration={markings, mark=at position 0.8 with {\arrow{stealth reversed};}}] (0,0) -- (2.05,0);
        
        \draw[orange, thick, postaction={decorate},decoration={markings, mark=at position 0.51 with {\arrow{stealth};}}] (-2.05,-0.48) -- (2.05,-0.48);
        \draw[orange, thick, postaction={decorate},decoration={markings, mark=at position 0.51 with {\arrow{stealth};}}] (-2.05,-0.62) -- (2.05,-0.62);
        \draw[blue, thick, postaction={decorate},decoration={markings, mark=at position 0.5 with {\arrow{stealth reversed};}}] (-2.05,-0.96) -- (2.05,-0.96);
        \draw[blue, thick, postaction={decorate},decoration={markings, mark=at position 0.5 with {\arrow{stealth reversed};}}] (-2.05,-1.1) -- (2.05,-1.1);
        \draw[blue, thick, postaction={decorate},decoration={markings, mark=at position 0.5 with {\arrow{stealth reversed};}}] (-2.05,-1.24) -- (2.05,-1.24);
        
    \end{tikzpicture}
    \caption{An example of a site configuration. Incoming flux is given by $P=3$, $Q=3$, while outgoing is $P=2$, $Q=4$, as can be seen by the number of thick and dashed lines. The change at the site is defined by the fermionic numbers.}
\end{figure}

One major concern is the error that can build up in quantum computation. A simple way to detect any single-bit-flip error is by noticing a violation of the Abelian Gauss law. The Hamiltonian commutes with the generators of the Gauss law. In this paper the oracle is designed, which checks whether this law is broken between two given neighbouring sites. The form of the law is as follows. Since all SU(3) indices are contracted in the LSH formulation by definition, there are no longer non-Abelian Gauss laws. Instead, there is the simple requirement that the amount of flux lines (of each type) coming out of a site match those that enter the next site. Both the number of flux lines passing through the site, given by the bosonic number operators ${n}_P$ and ${n}_Q$, and the string ends created on the site by its fermionic content, contribute to the total amount of flux lines. For the left $\underline 1$ and right $1$ side of each site, these are defined as
\begin{equation}\label{LSHflux}
\begin{aligned}
{P}(\underline{1}, r)&={n}_{P}(r)+{\nu}_{0}(r)\left(1-{\nu}_{1}(r)\right),\\
{Q}(\underline{1}, r)&={n}_{Q}(r)+{\nu}_{\underline{1}}(r)\left(1-{\nu}_{0}(r)\right), \\
{P}(1, r)&={n}_{P}(r)+{\nu}_{1}(r)\left(1-{\nu}_{0}(r)\right),\\
{Q}(1, r)&={n}_{Q}(r)+{\nu}_{0}(r)\left(1-{\nu}_{\underline{1}}(r)\right) .
\end{aligned}
\end{equation}
These numbers define the IRREP of the link on either side of the site, and can be related to the SU(3) Casimirs. Now the Abelian Gauss law has the simple form
\begin{equation}\label{AGL}
\begin{aligned}
P(\underline1,r+1)&=P(1,r),\\
Q(\underline1,r+1)&=Q(1,r).
\end{aligned}
\end{equation}

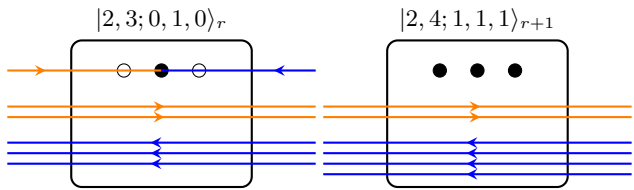
\begin{figure}[h]
    \centering
    \begin{tikzpicture}
        \draw[rounded corners, thick] (-1.2, -1.55) rectangle (1.2, 0.4);
        
        \node[above] at (0, 0.4) {$\ket{2,3;0,1,0}_r$};

        \draw (-0.5,0) circle (2.5pt);
        \filldraw (0,0) circle (2.5pt);
        \draw (0.5,0) circle (2.5pt);

        \draw[orange, thick, postaction={decorate},decoration={markings, mark=at position 0.25 with {\arrow{stealth};}}] (-2.05,0) -- (0,0);
        \draw[blue, thick, postaction={decorate},decoration={markings, mark=at position 0.8 with {\arrow{stealth reversed};}}] (0,0) -- (2.05,0);
        
        \draw[orange, thick, postaction={decorate},decoration={markings, mark=at position 0.51 with {\arrow{stealth};}}] (-2.05,-0.48) -- (2.05,-0.48);
        \draw[orange, thick, postaction={decorate},decoration={markings, mark=at position 0.51 with {\arrow{stealth};}}] (-2.05,-0.62) -- (2.05,-0.62);
        \draw[blue, thick, postaction={decorate},decoration={markings, mark=at position 0.5 with {\arrow{stealth reversed};}}] (-2.05,-0.96) -- (2.05,-0.96);
        \draw[blue, thick, postaction={decorate},decoration={markings, mark=at position 0.5 with {\arrow{stealth reversed};}}] (-2.05,-1.1) -- (2.05,-1.1);
        \draw[blue, thick, postaction={decorate},decoration={markings, mark=at position 0.5 with {\arrow{stealth reversed};}}] (-2.05,-1.24) -- (2.05,-1.24);

        \draw[rounded corners, thick] (3, -1.55) rectangle (5.4, 0.4);
        
        \node[above] at (4.2, 0.4) {$\ket{2,4;1,1,1}_{r+1}$};

        \filldraw (3.7,0) circle (2.5pt);
        \filldraw (4.2,0) circle (2.5pt);
        \filldraw (4.7,0) circle (2.5pt);
        
        \draw[orange, thick, postaction={decorate},decoration={markings, mark=at position 0.51 with {\arrow{stealth};}}] (2.15,-0.48) -- (6.25,-0.48);
        \draw[orange, thick, postaction={decorate},decoration={markings, mark=at position 0.51 with {\arrow{stealth};}}] (2.15,-0.62) -- (6.25,-0.62);
        \draw[blue, thick, postaction={decorate},decoration={markings, mark=at position 0.5 with {\arrow{stealth reversed};}}] (2.15,-0.96) -- (6.25,-0.96);
        \draw[blue, thick, postaction={decorate},decoration={markings, mark=at position 0.5 with {\arrow{stealth reversed};}}] (2.15,-1.1) -- (6.25,-1.1);
        \draw[blue, thick, postaction={decorate},decoration={markings, mark=at position 0.5 with {\arrow{stealth reversed};}}] (2.15,-1.24) -- (6.25,-1.24);
        \draw[blue, thick, postaction={decorate},decoration={markings, mark=at position 0.5 with {\arrow{stealth reversed};}}] (2.15,-1.38) -- (6.25,-1.38);

    \end{tikzpicture}
    \caption{Example of two neighbouring sites satisfying the non-Abelian Gauss law. The extra $Q$-type flux line generated by the first site simply passes through the second as a contribution to its $n_Q$, which is larger by one than the first site's. The second site also has a hadron (all three fermionic numbers are 1).}
    \label{nglflux}
\end{figure}

The AGL is never broken by Hamiltonian evolution, which is in agreement with the idea of having a physical sector of the entire Hilbert space. However, this Hilbert space is already much smaller than the original one from before contracting all SU(3) indices. What matters here is that if time evolution starts from a physical state, the AGL should keep being satisfied between every pair of neighbouring sites $(r,r+1)$. Any bit flip, as will be clear in the next section, will cause one of the quantum numbers to change. The resulting state will always violate the AGL if the original didn't. This can be deduced from (\ref{LSHflux}) and (\ref{AGL}). Thus, the whole class of single-bit errors during quantum computation can be detected by inspecting the AGL at all links. A second bit flip can cancel out the AGL violation of the first one (still constituting an error), but in most cases it will not. The same is true for further bit flips. In conclusion, most errors in general will also be detected by checking for AGL at all links.

To recapitulate, the resulting formulation, equivalent to the original Kogut-Susskind Hamiltonian, consists of a Hilbert space spanned by occupational quantum numbers local only to the lattice sites, and a Hamiltonian fully expressible in terms of their ladder (and number) operators. There are two bosonic quantum numbers due to the $SU(3)$ being a rank two group, and three fermionic numbers. These together define the electric flux configurations on either side of the site and the fermionic content at the site. The Abelian Gauss law enforces the total amount of each type of flux to match between neighbouring sites.

\section{Mapping to qubits}
\label{sec:Qubit}

An essential part of quantum simulation is the mapping of lattice variables onto qubits of a quantum computer. Such a system of qubits can then be used by algorithms for state preparation and time evolution, along with possible auxiliary qubits for calculations. For the simulated state to be physical, it needs to satisfy the AGL. One can initiate the qubits in a state that satisfies the laws, and then check for errors by probing it again during the evolution. This is the task of the oracle presented in the next section. In this section, the qubit number requirement for encoding the lattice variables is analyzed to show the advantage of LSH compared to the original formulation.

\subsection{irrep basis mapped to qubits}

In the Kogut-Susskind formulation there are $3$ fermionic numbers per site, just like in LSH, but the bosonic numbers are defined on links instead of the sites. For each link one needs to specify the irreducible representation (the same one for both sides of the link) and the basis state within it on each side. This forms an IRREP (electric) basis for the gauge field, just like $\ket{j,m,n}$ in the case of SU(2). The IRREP labels alone are not enough because the full basis states are needed to construct the nonlocal gauge-invariant states. The Gauss laws remain non-commuting and fully unsolved as well, but the number of qubits for encoding the state is significantly larger itself. A state of the link connecting sites $r$ and $r+1$ can be written
\begin{equation}
\ket{p,q;I_L,I^z_L,Y_L,I_R,I^z_R,Y_R}_r,
\end{equation}
where $p$ and $q$ label the IRREP, $I$ is the isospin, $I_z$ one of its projections and $Y$ the hypercharge. $p$ and $q$ are the numbers of upper and lower indices of tensors transforming in that IRREP. The numbers $P$ and $Q$ of (\ref{LSHflux}) are the numbers of columns with a single box and the number of columns with two boxes in the Young tableau, respectively. They can be used to label the IRREP as well. To have equivalence in the cutoffs, $P$ and $Q$ values are limited to $2^N-1$, meaning $N$ qubits are used to encode each using a simple binary base form of the numbers. This is not strictly equivalent since in the LSH encoding, we can have fluxes of $2^N$ on a link if there is a string end at an adjacent site, along with the maximum bosonic number. However, it is the closest they can get. The relation between the two sets of Casimirs is
\begin{equation}
\begin{aligned}
p&=P+Q\\
q&=P.
\end{aligned}
\end{equation}

\begin{figure*}[]
    \centering
    \subfloat[\label{qubit.cost.2}]{%
      \includegraphics[width=0.48\textwidth]{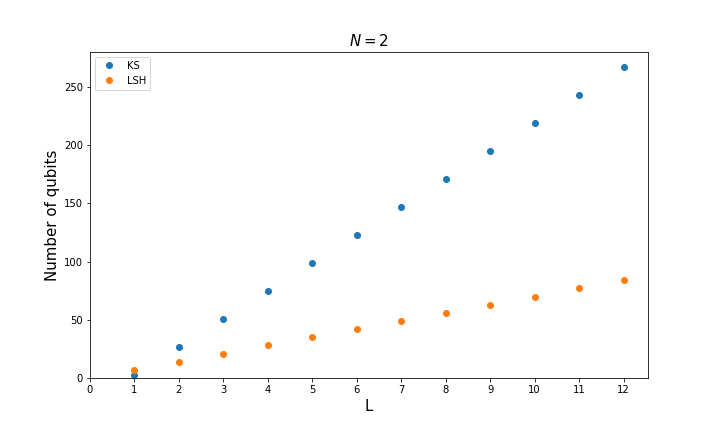}
    }
    \hfill
    \subfloat[\label{qubit.cost.5}]{%
      \includegraphics[width=0.48\textwidth]{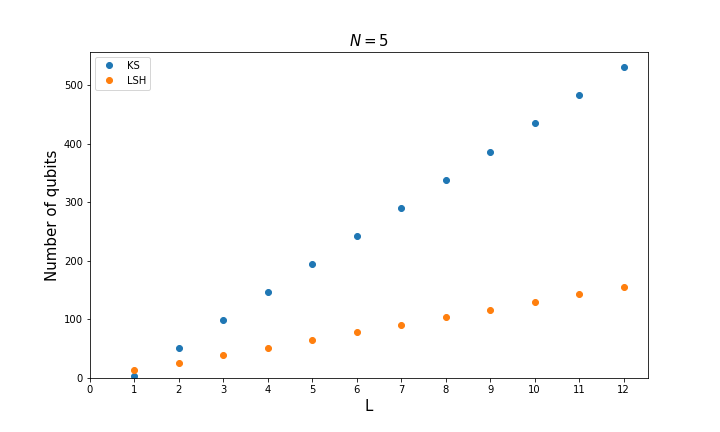}
    }
    \caption{The qubit cost of encoding the state of SU(3) LGT with fermions on a 1-dimensional lattice as a function of the number of sites $L$. Qubit digitization of the LSH formulation is compared to that of the Kogut-Susskind in terms of the number of qubits required to map the dynamic variables. A lower bound is used for the plot points in the Kogut-Susskind case. Two cases of the chromoelectric fields' cutoff are shown, where $N$ is the number of qubits storing each. Open boundary conditions are assumed, whereas other ones might add a constant term to both approaches, not affecting the scaling, as both methods use local variables.}
    \label{qubit.cost}
  \end{figure*}

Next, the matrix entry position specified by $I$, $I_z$, and $Y$ values on the two sides of the link also needs to be encoded, as it defines a basis state in the IRREP, and thus also in the Hilbert space. The number of matrix entries is given by the square of the dimension of the IRREP,
\begin{equation}
\begin{aligned}
d&=\frac{1}{2}(p+1)(q+1)(p+q+2)\\
&=\frac{1}{2}(P+Q+1)(P+1)(2P+Q+2).
\end{aligned}
\end{equation}
As the number of qubits in a simulation is constant and cannot depend on the representation, it has to be chosen based on the largest allowed values of $P$ and $Q$. As was noted above, $N$ qubits are used for encoding $P$ and $Q$ each, making their cutoff equal $2^N-1$, and the maximal dimension is
\begin{equation}
\begin{aligned}
d_{max}&=\frac{1}{2}(2(2^N-1)+1)(2^N-1+1)(3(2^N-1)+2)\\
&=3\cdot2^{3N+1}-5\cdot2^{2N}+2^N.
\end{aligned}
\end{equation}
The maximal number of states per IRREP is then $d_{max}^2$, and the number of qubits to encode them is
\begin{equation}
N_{states}=\lceil\log_2d_{max}^2\rceil,
\end{equation}
which is rounded up since the number of qubits is an integer and it cannot be less than required for the selected cutoff. For $N\geq2$ we have
\begin{equation}
\begin{aligned}
N_{states}&\geq6N+5.
\end{aligned}
\end{equation}
We conclude that in addition to the 3 qubits for fermion numbers per site, there are $\geq 8N+5$ qubits per link, where the $2N$ qubits for $P$ and $Q$ have been added to $N_{states}$. This is compared to the LSH cost in Figure \ref{qubit.cost}. Both clearly scale linearly in lattice size, but LSH with a much smaller slope.
\begin{equation}
n_{\textrm{qubits}}^{KS}=(8N+5)n_{\textrm{links}}+3n_{\textrm{sites}}
\end{equation}

\subsection{LSH degrees of freedom mapped to qubits}

Using the LSH framework, a 1-dimensional system of quarks and SU(3) gauge fields can be described by 5 quantum numbers per site. These define the LSH basis of states. After choosing a cutoff value, this basis can be directly mapped onto qubits of a quantum computer. Each of the three fermionic numbers is represented by a single qubit. The two bosonic numbers are, in theory, unbounded, so the simulation has to be limited to a subspace defined by a chromoelectric fields' cutoff value ${\Lambda}$. This corresponds to specifying the maximum value for the bosonic quantum numbers. Thus, each of them is encoded by a finite number $N$ of qubits, which can be chosen based on the available resources of the hardware being used. The bosonic numbers are integers starting from zero, so they are mapped onto the qubits as binary numbers. The possibility of this qubit digitization is one of the benefits of the LSH framework, as the number of qubits is greatly reduced by using this basis. The total number of qubits per site is $2N+3$.

Current quantum computers are often severely limited by their connectivity, which constrains the interactions qubits can have with each other via quantum gates. Connectivity is a function of the computer's architecture, which must be balanced for coherence, fault tolerance, and other factors. Here, the design's viability is not considered for any real quantum hardware; instead, universal connectivity is assumed. In other words, for the exact qubit digitization method (eg, the order of qubits) and implementability of the algorithm that will be given in the next section, one will need to consider appropriate architecture and its details, as well as gate availability. This work focuses on the algorithm's logic.

To be able to compare with the resource requirements of other methods, the details of how the cutoffs work need to be considered. As Equations (\ref{LSHflux}) show, sites can create flux to their sides (the second term in each expression's right-hand side), which is added up with the amount of flux already passing through the site ($n_P$ or $n_Q$). If the next site doesn't immediately absorb this additional flux line, it will be incorporated into the bosonic number $n_P$ or $n_Q$ at that site. Visually, this translates to a site creating flux that passes through the next site, as in Figure \ref{nglflux}. Suppose the first site already had the bosonic number at the maximum value for the chosen number of qubits representing it. Then the next site will have that number overflown. However, if the next site immediately absorbs this flux line, there is no overflow. The fact that $N$ qubits are used per bosonic number (in a binary number encoding fashion) therefore limits the bosonic numbers to $2^N-1$, though the flux itself can be equal to $2^N$ if its string-ends are at neighbouring sites. Thus, the fluxes are limited to
\begin{equation}
\Lambda=2^N,
\end{equation}
as this is what the links see.

In conclusion, since all variables in LSH are on-site, we simply have $2N+3$ qubits per site, $N$ for each bosonic and $1$ for each fermionic number. This gives a field cutoff of $2^N$.
\begin{equation}
n_{\textrm{qubits}}^{LSH}=(2N+3)n_{\textrm{sites}}
\end{equation}

\section{Oracle to check gauge invariance}
\label{sec:oracle}

\begin{figure*}[]
    \centering
    \includegraphics[scale=0.74]{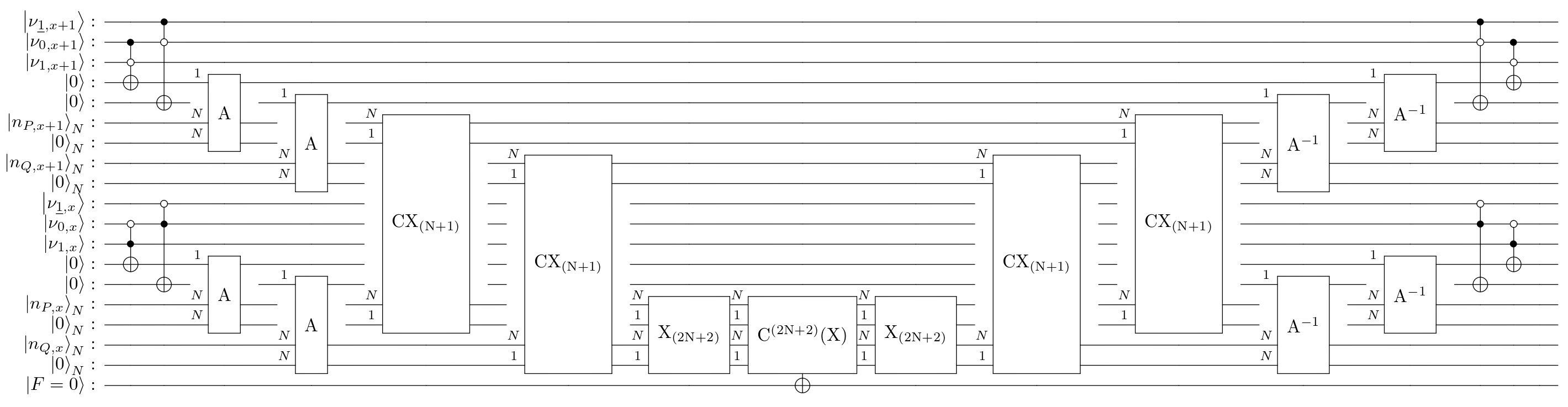}
    \caption{{The oracle which computes the Abelian Gauss law and puts the result into the flag qubit $\ket F$. The design is heavily inspired by a previous work on the SU(2) version of the theory \cite{Gauss}. The main difference is in additional second layers of adders and comparators, a consequence of the fact that SU(3) is a rank 2 group.}}
    \label{oracle}
\end{figure*}

The Hilbert space that is spanned by the LSH quantum numbers encoded in qubits includes all possible combinations of fluxes going through and being generated at each site of the lattice. Within it, only the states for which the amounts of flux are consistent between each pair of neighbouring sites are physical. In other words, the physical subspace of the space spanned by the qubits is constrained by the Abelian Gauss law. The idea behind this paper is that the AGL can be phrased as an operator acting on the system's qubits, and this operator can be decomposed into quantum gates. The result of the operator is binary - either the law is satisfied or it isn't. The oracle presented here (Figure \ref{oracle}) takes in the qubits of two adjacent sites, along with some auxillary qubits, evaluates the numbers of flux lines of each type leaving one site and those entering the other, checks if they are the same as per the AGL, and leaves a designated qubit in one of its basis states based on the result. The rest of the qubits are then reverted to their original state. The flagged qubit holds the information about whether the AGL is broken or not. It can then be used in further computation to alter the rest of the simulation or discard the results when the law was violated.

\subsection{Algorithm}
\label{sec:alg}

The overall structure of the oracle consists of a computation of the two sides of the AGL using qubits of two neighbouring sites and some ancillas, a comparison, a controlled change of the flag qubit's state, and a reverse computation of all qubits but the flag. The first part uses double-controlled NOT gates where one of the controls is activated when the qubit is in state 0 and not 1 (which is easily implemented as a Toffoli gate with an X (NOT) gate before and after it on that qubit) for the fermionic terms, and adder subcircuits (labelled $\textrm A$) for adding them to the bosonic terms (see \ref{LSHflux}). The results for each type of flux $P$ and $Q$ are compared (cf. \ref{AGL}) by applying CNOT gates pair-wise to all qubits representing the two calculated values in each of the equations. These are labelled $\textrm{CX}_{\textrm{(N+1)}}$ as they represent $N+1$ CNOT (CX) gates, coming from $N+1$ qubits encoding each site's flux number. There are two such sets of CNOT gates, called comparators, one for each type of flux. The additional qubit beyond the $N$ qubits for encoding $n_P$ and $n_Q$ is the final carry bit from the addition, and will be zero if the state is within the cutoff. The oracle, however, doesn't see this and simply compares the numbers even if the addition has made them overflow the cutoff (alternatively, one could measure the carry bit to see if the state is within the cutoff, or one could assume that it is, and not even compute the last carry digit, allowing for 1 fewer ancilla qubit per bosonic number in the oracle). At this point in the circuit, all of the results of the CNOTs are in state 0 iff the AGLs are satisfied. The set of X gates that come next, $\textrm{X}_{\textrm{(2N+2)}}$, flips all of them. The controlled X gate $\textrm{C}^{\textrm{(2N+2)}}\textrm X$ flips the state of the flag qubit $\ket F$ iff all of the control qubits are in state 1 (which is now the case iff the AGLs are satisfied). The rest of the circuit applies all of the gates in reverse, except for $\textrm{C}^{\textrm{(2N+2)}}\textrm X$.

Since the $\textrm{C}^{\textrm{(2N+2)}}\textrm X$ gate can only change the flag qubit $\ket F$, the reverse computation returns all other qubits to their original states and entanglement. However, the state of $\ket F$ is flipped into $\ket{F=1}$ in the parts of the overall superposition for which the AGLs are satisfied. The state $\ket{F=0}$, on the other hand, becomes entangled with the parts where at least one of the AGLs is broken.

\begin{figure}[]
    \includegraphics[scale=0.4]{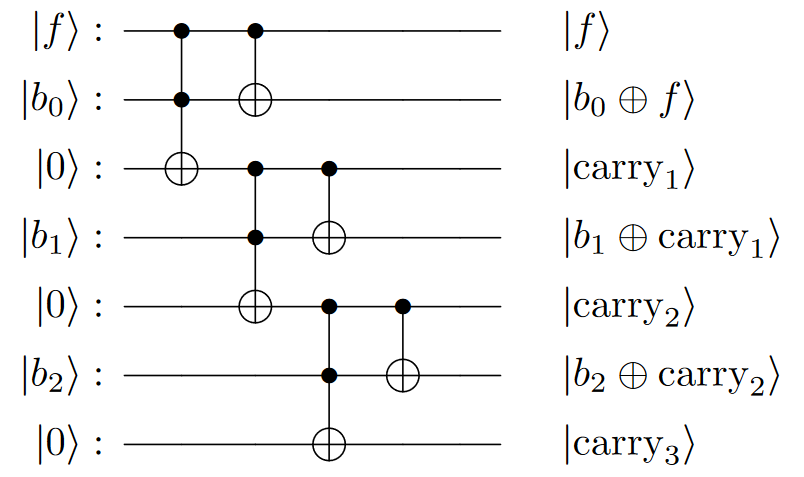}
    \caption{Adder subcircuit for $N=3$. The fermionic term's qubit $\ket{f}$ is added to the bosonic number's qubits. The carry bits are computed along the way, and the last carry bit is the overflow-bit in the result. The subcircuit uses $N$ ancilla qubits, $N$ CNOT gates and $N$ Toffoli gates. $A^{-1}$ is its inverse, applying the same gates in the opposite direction.}
    \label{adder}
\end{figure}

For the purpose of resource requirement calculations, the adder is assumed to be implemented as a simple addition of a one-qubit number to a binary number represented by $N$ qubits, with $N$ ancillas used for the carry bits. It uses Toffoli gates for carry calculation and CNOTs for the bit-wise addition, as shown in Figure \ref{adder} for the case $N=3$. The result is located in the $N$ qubits of the original $N$-qubit number and the additional last carry qubit. These are then the inputs for a comparator subcircuit.

\subsection{Component requirement}
\label{sec:comp}

For testing on hardware, it is of utmost importance to have as few qubits, gates, and most notably sequential layers of gates as possible. The smaller these numbers are, the less noise and errors there will be. Gates and layer depth are evaluated for a circuit's decomposition into single-qubit and two-qubit gates. The more important among these are the 2-qubit gates, as that's where entanglement is introduced and the risk of environment interference is highest. Existing quantum computers usually come with a set of gates that can be applied directly, while other gates will first have to be expressed (or approximated) as their combinations. The number of two-qubit gates and their layers is usually independent of the hardware, though, as going from one set to another will only add single-qubit gates.

In the present case, the required resources depend on the size of the simulated system and on the cutoff value $\Lambda$ of the chromoelectric fields. The lattice size sets the number of sites and oracles, while the cutoff value determines the number of qubits per site and the number of quantum gates used to implement the subcircuits. One can also change the implementation of the adder subcircuits used, possibly changing the number of ancilla qubits. As mentioned, it is assumed here that adders are implemented as in Figure \ref{adder}, while the multi-controlled NOT gate is decomposed into Toffolis using lemma 7.2 of \cite{Gates}.

Just as in \cite{Gauss}, the algorithm doesn't actually require exact Toffoli gates most of the time. They only need to create outputs that activate the multi-controlled CX, which allows for gates that differ from exact Toffolis by phases introduced on individual qubits. The controlled gates get applied regardless of the phases, as long as the controls as in the state $\ket1$ each. These phases get uncomputed in the second half of the circuit by using the same implementation in reverse. Such an operation is called a Toffoli-congruent gate and can be implemented by using only 3 CNOT gates (and 4 single-qubit gates) \cite{Gates}. The resource counting here assumes these to be used in the adders and also as the eight explicit Toffolis in the circuit. Toffolis in the decomposition of the multi-controlled CX gate have to be exact, however, since this gate does not get uncomputed and it must not introduce relative phases in the overall state. They are implemented by 6 CNOTs (and 8 single-qubit gates), as per Corollary 6.2 in \cite{Gates}. These results are used to compute the CNOT equivalent in table \ref{gates}.

\begin{table}[]
\begin{tabular}{l|l}
Resource               & Count \\ \hline
Site state qubits      & $2N+3$  \\
Ancillas per site      & $2N+2$  \\
Total qubits in an oracle & $8N+11$ \\
Sites                  & $L$     \\
Oracles (= links)      & $L-1$   \\
Adders per oracle      & $8$     \\
Multi-C X's per oracle & $1$     \\

\end{tabular}
\caption{Resources for encoding the state and applying oracles to all links.}
\label{counts}
\end{table}

\begin{table}[]
\begin{tabular}{l|l|l|l}
Counts of gates     & CNOT  & Toffoli & CNOT equiv. \\ \hline
Adder               & $N$     & $N$       & $4N$          \\
Multi-controlled X  & $0$     & $4N-8$    & $24N-48$      \\
Other parts         & $4N+4$  & $8$       & $4N+28$       \\
Oracle              & $12N+4$ & $12N$     & $60N-20$     
\end{tabular}
\caption{Gate counts for subcircuits and the oracle assuming $N\geq3$.}
\label{gates}
\end{table}

\subsection{Depth}
\label{sec:depth}

The final number to calculate in this analysis is the depth of the circuit in terms of parallelizable layers of gates. Quantum computers have a limited amount of time in which the errors are managable since decoherence due to noise grows in time, which is determined by the number of parallel steps of operations. The physical depth, realized by an implementation on specific hardware, will depend on the connectivity of the physical qubits and gates. Two-qubit gates can usually not be applied to physically distant qubits, and the oracle does not allow for a simple construction with all the interacting qubits neighbouring each other. Thus, there will likely be some swap gates applied in between certain subcircuits to get the qubits in place, adding to the depth. However, this depends on the implementation, so here we focus on logical depth, accounting only for the desired operations themselves.

As two-qubit gates usually take significantly longer time, and additional single-qubit gates might be necessary to convert the circuit into gates available for the hardware, only two-qubit gates will be the focus here. In particular, the CNOTs counted in the previous subsection. The explicit Toffoli-congruent gates at the beginning and end of the oracle constitute three consecutive CNOTs each. Looking at this implementation in subsection VI.B. of \cite{Gates}, one can see that the two layers of these Toffolis at each side of the oracle can be run in parallel despite sharing one of the control qubits. The adders' CNOTs can be run in parallel with their Toffoli-congruents, except for the last one (see Figure \ref{adder}). The four adders themselves all operate on different sets of qubits and can thus run in parallel. The same is true of the two comparator $(\textrm{CX}_{\textrm{(N+1)}})$ gates, which are therefore a single layer of CNOTs. The multi-controlled $\textrm C^{(2N+2)}(\textrm X)$ has the greatest depth cost, equaling the total number of CNOTs in its decomposition. The depths are computed using corollary 6.2 and lemma 7.2 from \cite{Gates} and shown in table \ref{depth}.

\begin{table}[h]
\begin{tabular}{l|l}
Circuit                & CNOT depth \\ \hline
Explicit Toffolis      & $2\times3$  \\
Adders      & $2\times(3N+1)$  \\
Comparators      & $2\times1$  \\
Multi-controlled NOT      & $24N-48$  \\
Oracle total      & $30N-38$  \\

\end{tabular}
\caption{Depth in terms of the number of two-qubit (CNOT) gate layers is shown for subcircuits and the oracle. Multiplication by $2$ is left explicit as it comes from having the inverses of most parts in the second half of the oracle. $N\geq3$ is assumed.}
\label{depth}
\end{table}

\section{Mapping to qudits: a future prospect}
\label{sec:Qudit}

Instead of a qubit, one can use a single multi-level quantum computer cell (a qudit) to represent the state of an integer quantum number. If the chromoelectric fluxes were encoded this way, the cutoff would be defined by the number of levels (basis states) of the qudit. For example, a 3-level qudit can represent flux number values from 0 to 2. However, the hardware would have to be completely changed for each increase in the cutoff that one might want to realize. A more practical utilization of qudits is to simply use them the same way as qubits. Namely, instead of using binary numbers as with qubits, one can store the integer quantum numbers in base $d$, where $d$ is the number of levels of each qudit.

This entails the usage of quantum gates appropriate to the base $d$. An example is the increase of a bosonic number by one. It would be implemented by a $d$-base equivalent of the adder algorithm \ref{adder}. In this version, each qudit can represent a number from $0$ to $d-1$, and each digit in a multi-qudit register represents this number multiplied by $d^n$, where $n$ is the qudit's position in the register (starting from zero). For example, in base $4$ the first qubit's state represents numbers $0$ to $3$, the second one's is multiplied by $4$, the third's by $16$ etc. Instead of the CNOT determining a sum of two bits, one now needs to raise the state of one qudit the number of levels equal to the state of the other. If this takes the qudit over the state $d-1$, it should overflow and continue from state $0$. If this happens, a carry qudit (or qubit) should be flagged and should add one to the next digit's addition. All of these operations are simple unitary operators, just like in the qubit case, and map onto the standard qudit gates.

Apart from these adaptations at the subcircuit level, our oracle would schematically look the same in the qudit version, and it would have the same output and usage possibilities. It could also keep using qubits for the fermionic numbers and the flag bit, if the architecture can use both qubits and qudits.

Qudit systems have been studied in a number of works recently, with algorithms being developed and various hardware realizations considered \cite{Wang2020Qudits}. For instance, photonic integrated chips \cite{Chi:2022uql,Chi:2023xtt} and superconducting quantum computers \cite{Fischer:2022dbr,Tripathi:2024wom} are considered along with gates, general algorithms and error mitigation techniques such as dynamical decoupling. Qudits have been considered in lattice gauge theory development for simulations \cite{Ciavarella:2021nmj}, and work has been done on simulations using a trapped-ion qudit quantum processor \cite{Meth:2023wzd}.

The main advantage of using qudits alongside qubits is that the numbers of qudits, gates, and gate layers required for the simulation are reduced. The same amount of physical components (qudits) gives a larger computational space than with qubits. The bosonic numbers are encoded into $\lceil\log_d(\Lambda+1)\rceil$ qudits, and the adders, comparators, and the multi-controlled NOT gate all operate on fewer qudits, reducing the needed amount of gates and layers as well. This is most useful regarding the goal of increasing the cutoff on the fluxes as much as possible with a limited amount of resources and short coherence time.

\section{Discussions and outlook}
\label{sec:discus}

Qubit digitization of SU(3) lattice gauge theory with fermions on a 1-dimensional lattice was laid out using the Loop-String-Hadron framework. It is clearly more efficient than naive Kogut-Susskind digitization, and it leaves only the Abelian simplification of the Gauss law to be checked. The proposed oracle for checking the laws on each link was explained, and its qubit, gate, and depth costs were analyzed.

Hopefully, the oracle can be tested on quantum hardware in the near future, as it should eventually serve as a tool for error correction or a criterion for discarding results as unphysical. For instance, it can be a part of a larger circuit simulating QCD dynamics. This circuit would consist of the time evolution algorithm (which is in development at the time of writing), as well as other error mitigation modules.

Another direction actively being pursued is the development of a consistent description of higher-dimensional lattices in terms of LSH degrees of freedom. In lower dimensions, one can hope to see some of the phenomena intrinsic to quarks and the strong force, but the ultimate objective is the simulation of lattice QCD in its full form.

\section{Acknowledgments}
The authors would like to thank Himadri Mukherjee and Jesse Stryker for insightful conversations at various points throughout this work and for sharing comments on the manuscript.
Research of IR is supported by the  OPERA award (FR/SCM/11-Dec-2020/PHY) from BITS-Pilani, the Start-up Research Grant (SRG/2022/000972) and Core-Research Grant (CRG/2022/007312) from ANRF, India and the cross-discipline research fund (C1/23/185) from BITS-Pilani. FI is supported by the cross-discipline research fund (C1/23/185) from BITS-Pilani, received by IR. The authors acknowledge useful discussions during meetings of QC4HEP working group.

\bibliography{bibi.bib}

\end{document}